\documentclass[sigconf]{acmart}

\usepackage{amsmath,amsfonts}
\usepackage{graphicx}
\usepackage{textcomp}
\usepackage{xcolor}
\usepackage{xspace}
\usepackage{epstopdf}
\usepackage{url}
\usepackage{natbib}

\usepackage{booktabs}
\usepackage{listings}

\usepackage{xfp}

\setcopyright{acmcopyright}
\copyrightyear{2022}
\acmYear{2022}
\acmDOI{XXXXXXX.XXXXXXX}

\acmConference[SE 2030]{International Workshop on Software Engineering in 2030}{November 2024}{Puerto Galinàs (Brazil)}
\acmPrice{15.00}
\acmISBN{978-1-4503-XXXX-X/18/06}

\begin{document}
	
	\title{Position: How Regulation Will Change Software Security Research}
	\newcommand{\todo}[1]{{\textcolor{red}{\textbf{TODO: }#1}}\xspace}
\newcommand{\toolname}{\textsc{ValBench}\xspace}

\lstset{language=Java,
	stringstyle=\ttfamily,
	basicstyle=\ttfamily,
	columns=fullflexible,
	numbers=left,
	xleftmargin=2em,
	escapechar={$},
	breaklines=true
}

\newcommand*\circled[1]{\tikz[baseline=(char.base)]{
		\node[shape=circle,draw,inner sep=2pt] (char) {#1};}}

\setlength{\textfloatsep}{0.1cm}
\setlength{\floatsep}{0.1cm}

	\author{Steven Arzt}
	\email{steven.arzt@sit.fraunhofer.de}
	\orcid{0000-0002-5807-9431}
	\author{Linda Schreiber}
	\email{linda.schreiber@sit.fraunhofer.de}
	\author{Dominik Appelt}
	\email{dominik.appelt@sit.fraunhofer.de}
	\affiliation{%
		\institution{Fraunhofer SIT | ATHENE}
		\streetaddress{Rheinstraße 75}
		\city{Darmstadt}
		\state{Hessen}
		\country{Germany}
		\postcode{64295}
	}
	
	\begin{abstract}
		Software security has been an important research topic over the years. The community has proposed processes and tools for secure software development and security analysis. However, a significant number of vulnerabilities remains in real-world software-driven systems and products.

To alleviate this problem, legislation is being established to oblige manufacturers, for example, to comply with essential security requirements and to establish appropriate development practices. We argue that software engineering research needs to provide better tools and support that helps industry comply with the new standards while retaining efficient processes. We argue for a stronger cooperation between legal scholars and computer scientists, and for bridging the gap between higher-level regulation and code-level engineering.

	\end{abstract}


	
	\maketitle

	\section{Introduction}
	\label{sec:intro}
	With the increasing digitialization of every part of life, IT security has become a core topic in software engineering. It is long known that mobile devices, for example, are used to perform sensitive operations such as transferring money or unlocking doors. The apps running on these devices have access to sensor data, e.g., GPS positions, but also calendar entries, text messages, e-health records,  and similar sensitive data. These security challenges are, however, no longer specific to the mobile ecosystem. Smart features are nowadays part of many products, from TV sets to refrigerators, from cars to insuline pumps. All of these systems are interconnected and thereby provide an attack surface with which attackers can interact remotely. Successful attacks can have severe consequences, ranging from surveillance and data theft in the most private parts of life to physically endangering lifes.

These challenges are known to the research community. Existing work on software security has explored constructive approaches such as more secure programming languages~\cite{10.1145/3158154,safelang}, provably secure systems~\cite{lortz2014cassandra} and formal methods~\cite{de2019efficient,ball2004slam}, risk modeling techniques~\cite{agence2019ebios,iso27005} as well as static~\cite{arzt2014flowdroid,livshits2005finding,jovanovic2006pixy} and dynamic~\cite{enck2014taintdroid} code analysis techniques for identifying vulnerabilities. Policy enforcement~\cite{rasthofer2014droidforce} has been researched as a technique for reducing the impact of a security issue. When new programming models such as \emph{mini apps}~\cite{wang2023uncovering,10.1145/3605762.3624432,10409279} or \emph{smart contracts}~\cite{wang2021ethereum,10.1145/3385412.3385990,10.1145/3243734.3243780} are invented, the community has reacted with security analyses for them.

Despite these efforts, the number of CVEs published per year continues to grow. Similarly, it has been shown that the total number of security-related code smells in Android apps \emph{increases} over the years~\cite{arzt2022security}. This is expected as the number, size and complexity of IT systems grows, but also makes a society and the individual less resilient against cyber attacks. Further, the level of security varies greatly between products. While some manufacturers operate dedicated security teams and follow strict secure software engineering processes, others fall short of such organizational measures. Without suitable processes, code quality and security may suffer accordingly~\cite{arzt2021security}. In total, the efforts of the IT security and software engineering communities have not lead to a sufficient level of security in the real world.

Regulators have therefore identified the need to enact binding rules on how data or systems must be protected. The \emph{Cyber Resilience Act} (CRA)~\cite{cra} of the European Union is the latest example of such regulation. The CRA provides a list of essential security requirements along with necessary organizational processes, e.g., for coordinated vulnerability disclosure. Further, the CRA establishes liabilities and penalties for manufacturers that fail to comply with these requirements.

In this paper, we argue that existing and upcoming regulatory efforts will lead to new stakeholders in software security research and industrial application. We highlight how compliance and security overlap and argue for a better integration between legal and engineering research. We are certain that regulation, along with its requirements and deadlines for the industry, will structure and prioritize future research challenges. The software industry must ensure compliance and academic research must provide the tools and techniques that link traditional software security research with what is legally required. This includes not only measures to be taken, but also how evidence for security measures must be given, and which guarantees are required from the finished product for avoiding fines and potentially liability. We argue for the need for regulation-compliant certification schemes that bridge the gap between high-level legal requirements and technical measures. At the same time, we argue that software security research should provide input to shape future legislation towards what is necessary, possible and practical.

In this paper, we focus on EU regulation. However, the case of privacy protection in IT systems has shown that, following the EU example, similar regulation is enacted in other countries as well~\cite{kober2023sensitive}. We therefore consider it possible that the EU's CRA can serve as a blueprint for similar legislation in other countries as well.

The remainder of this paper is structured as follows. We first provide some background on IT security regulation in Section~\ref{sec:background} before explaining the requirements of the EU's CRA and its impact on software engineering in more detail in Section~\ref{sec:requirements} and highlight the concrete research challenges that arise from the CRA in Section~\ref{sec:challenges} before concluding the paper in Section~\ref{sec:conclusion}.

	\section{Regulatory Background}
	\label{sec:background}
	In recent years, the number of (proposed) laws regulating how data and IT systems are to be used and protected has increased considerably. The \emph{General Data Protection Regulation} (GDPR)~\cite{EuropeanParliament2016a} defines rules on how personal data must be protected when processed. The \emph{Artificial Intelligence Act} (AI Act)~\cite{aiact} provides rules on e.g. the placing on the market, the putting into service, the use and transparency of artificial intelligence systems in the EU. The \emph{Digital Operational Resilience Act} (DORA)~\cite{dora} defines rules for the IT operations of financial entities. The \emph{Cyber Resilience Act} (CRA), which is the core focus of this paper, regulates products with digital elements to ensure the security of such products.

These regulations are just a few of the examples that have been introduced at EU level. As the regulated systems are mostly driven by software, the regulations pose a (rising) challenge on software engineering, especially in the context of IT security. They affect how software is designed, implemented, tested, released and maintained over its lifespan. Some regulations further create new demand for security software by, e.g., mandating intrusion detection systems (DORA) for regulated institutions.

This more recent legislation greatly broadens the scope of affected industries beyond these traditionally regulated industries. While manufacturers of medical devices or aircraft have experience with rigorous (safety) requirements and documentation, such processes must now be established in other industries as well. The CRA, for example, affects all products that are made available on the market in the course of a commercial activity. This broad definition can also apply to, e.g., free mobile gaming apps that generate revenue through in-app advertisement even if the app is provided by an individual developer.

 In the following, some key principles of regulation are introduced and discussed which can be found through various EU laws.

\textbf{Technological neutrality:} Regulations are intentionally designed to be technology-neutral, applicable to various technologies rather than specific ones (see Art. 2 GDPR, Art. 2 CRA, Art. 2 AI Act). Consequently, legal texts often remain very general, while the concretization and specification of the legal framework is up for discussion between different stakeholders from practice, science, jurisdiction and supervisory authorities. Therefore, the software research and industry communities must actively engage in this interdisciplinary exchange to incorporate its (justified) interests into the process of concretization and specification, in the best case by establishing (binding) specific standards as an outcome. Translating legal requirements into precise technical recommendations, instructions and standards ensures efficient compliance and prevents regulatory uncertainty from impeding technological progress. Some government agencies have issued standards, e.g., on the use of cryptography, e.g., BSI TR-02102-1 or ENISA's 2013 report on algorithms, key sizes and parameters. Still, these standards address low-level concerns that are easy to standardize regardless of the product at hand. Therefore, a large gap remains between the legal text and the product-specific implementation.

\textbf{Risk-Based Approach, Appropriateness, Proportionality:} Many regulations follow a risk-based approach, requiring only in very general terms that the measures to be taken must be adequate and/or proportionate in relation to the existing risk to be mitigated (see Art. 32 (1) GDPR, Annex I Part 1 CRA). However, the legal texts often lack guidance on conducting risk assessments and determining suitable measures. Again, this necessitates further specification through court judgments, guidance from supervisory authorities, standards from auditing organizations and more. The software industry must actively contribute its specific needs and expertise into this process. Furthermore, to determine which measures can be seen as adequate and/or proportionate to the risk requires a comprehensive risk assessment to be conducted first. The software industry thus faces the challenge of not viewing (software) components in isolation (e.g., in terms of their security) but always putting them in the overall context of the application. As the decision, which measures are adequate and/or proportionate according to the risk in general is at the discretion of the company concerned, the software industry has to develop comprehensible metrics and standards to quantify and weigh both the risk to be mitigated and the measures to be taken (according to the risk) against each other. As will be elaborated further below, a comprehensible documentation is the key to compliance.

\textbf{Privacy and Security by Design:} Legal requirements no longer relate solely to the finished (software) product, but increasingly to the entire software development and implementation process. For privacy and security requirements it is already mandatory to consider them from the very beginning in the design and development phase of the software (see Art. 25 GDPR, Art. 13 (1) CRA). In the future, it will therefore be essential that software engineers possess a fundamental understanding of legal principles and develop awareness of potential (legal) risks relating to their daily business, enabling them to seek appropriate legal guidance when necessary.

\textbf{Documentation obligations}: Comprehensive documentation of the measures implemented to meet legal requirements is essential (see Art. 13 (12) CRA, Art. 11 AI Act). This includes not only the need for documentation where there is an explicit legal obligation, rather comprehensive documentation should be provided wherever possible. As already pointed out above,  the risk-based approach always requires a case-by-case assessment of which specific measures can be considered as appropriate and/or proportionate. In general, it is at the discretion of the affected organization how to perform this assessment as well its final result. However, the result of this assessment can be subsequently questioned by a supervisory authority or a court. Therefore, the questioned appropriateness and/or proportionality will be almost impossible to prove without comprehensive documentation of the underlying considerations. The software industry is therefore faced with the challenge of finding mechanisms to efficiently document and provide evidence for decision-making processes (even where there is no legal obligation) including centralized storage of these. Transparency and accountability are not a question of what must be documented, but what can be documented, in a way which is comprehensible to third parties. At the same time, the impact on (agile) developmemt processes must be limited to retain economic viability.

The discussed points highlight the significant challenges that regulation can pose on software engineering. However, by actively engaging in the regulatory process, implementing robust risk assessment models, prioritizing privacy and security from the early stages of development, and maintaining comprehensive documentation, the software industry can effectively overcome these challenges. This proactive approach enables the industry to contribute to technological progress while adhering to legal requirements.

	\section{CRA Security Requirements}
	\label{sec:requirements}
	The following section focuses on analyzing the specific requirements arising from the Cyber Resilience Act as a current example of technology legislation by the EU.
Unlike the GDPR, which is centered on personal data, or other regulations that protect specific sectors, the CRA is intended to create a basic level of protection and therefore a security standard for all products, regardless of the areas in which they are used or the type of information they process (with few exemptions for products that are already subject to stricter cybersecurity legislation, such as aviation).

The CRA applies to \emph{products with digital elements} that are intended to have a direct or indirect logical or physical data connection to a device or network (Article 2, 1. CRA). Products with digital elements are defined as software or hardware products and their remote data processing solutions, including software or hardware components being placed on the market separately (Article 3(1) CRA). The inclusion of \emph{remote data processing solutions} is intended to ensure that products are covered in their entirety with all functions, regardless of whether data is therefore stored or processed locally on a user's device or remotely by the manufacturer (recital 11 CRA).
Manufacturers are obligated to design, develop and produce such products in such a way that they ensure ``an appropriate level of cybersecurity based on the risks'' (Annex I, Part I (1) CRA). With this risk-based approach, as described in section 2, manufacturers have the discretion to choose specific technical solutions or tools and thereby must take into account the individual risk factors of the product, such as e.g. the intended purpose or operational environment.

However, the CRA provides a more detailed level of specifications regarding the cybersecurity requirements to be implemented than, for example, the GDPR. These so-called \emph{essential requirements} are listed centrally in Annex I CRA. The following is a selection of  requirements listed in Annex I, Part I (2) CRA, that have to be implemented by manufacturers based on the results of the risk assessment:

\begin{description}
\item[Without known vulnerabilities] The product must be placed on the market without known \emph{exploitable} vulnerabilities. This requires a check for vulnerabilities in the product's own code, and in the dependencies.
\item[Secure by default configuration] Products must be made available on the market in a secure by default configuration with a possibility to reset the product to this secure initial state.
\item[Security updates] Manufacturers must ensure that vulnerabilities can be addressed through security updates, with automatic installation being the default from which users can opt out.
\item[Authorization] Products must be protected from unauthorized access with reporting of failed access attempts
\item[Confidentiality] Products must ensure the confidentiality of the data it handles using \emph{state of the art mechanisms}.
\item[Integrity] Products must protect the integrity of data during storage, processing and transport. This includes program files and configuration. The product must report on corruptions.
\item[Data minimization] Products may only use the data that is necessary for the intended purpose of the product
\item[Availability] Products must ensure th availability of basic and essential functions even in the presence of denial-of-service (DoS) attacks
\item[Minimize impact] Products must minimize the impact that a successful attack has on other connected devices
\item[Limit attack surface] Products must minimize their attack surface, including omitting unnecessary external interfaces
\end{description}

Furthermore, Annex I, Part II CRA requires manufacturers to put in place certain processes and measures relating to the handling of vulnerabilities, such as drawing up a software bill of materials (SBOM) ``in a commonly used and machine-readable format covering at the very least the top-level dependencies of the products'' or putting in place and enforcing a policy on coordinated vulnerability disclosure.

With requirements that already refer to the design phase of a product, the CRA poses a particular challenge for legacy software-based systems with long lifespans. While short-lived software such as mobile apps can be re-designed and re-implemented from scratch, this is not possible for complex industrial or backend software.  Technically, the CRA does not affect the provisioning of replacement parts on the market, but it applies to new versions of products with digital elements. In other words, a control unit of an existing industrial production machine may continue to be sold along with the software running on it. However, when a new version of the machine shall be put on the market, the new version must comply with \emph{all} requirements of the CRA, regardless of it being built on a significant legacy software (and hardware) stack.

Compliance with the requirements of the CRA must be demonstrated for each product through a \emph{conformity assessment procedure}. The CRA distinguishes between different categories of products with digital elements, depending on their criticality and importance (Article 7 and 8, Annex III and IV CRA). Based on this categorization, different procedures are available for the conformity assessment to choose from for  manufacturers, these are structured in modules and described in Article 32, Annex VIII CRA. Module A 'internal control procedure' is a self-assessment by the manufacturer; combined Module B and C are an 'EU-type examination procedure', followed by an 'internal production control' and module H is based on a 'full quality ensurance'. While module A can be carried out independently without the involvement of a third party, the other modules require the involvement of an external notified body.



	\section{Challenges}
	\label{sec:challenges}
	In this section, we focus some of the challenges that the CRA provides for software engineering research and industry. We acknowledge that some of these challenges were already (partially) present as research topics in the community. Regulation such as the CRA, however, requires a major shift in research priorities.

\paragraph{Compliance Standards}

Most software products only require the Module A conformity assessment procedure, i.e., internal checks by the manufacturer who then affixes the CE mark on their product without external validation. While convenient, this approach leaves all responsibility and liability with the manufacturer. The commumnity needs to work with market surveillance authorities to develop standards and procedures that operationalize the higher-level technical requirements from the CRA. While the CRA tasks the EU commission with providing guidance on certain types of products, the community should actively engage in the process and propose draft standards.

\paragraph{Product-Level Tooling}

Devising processes and tools for rooting design and security checks (code analysis, fuzzing, etc.) in a risk model is one of the challenges that the community faces. The requirements of the CRA are not component-specific, but address the entire product. Especially code scanners are, at the moment, oten component-specific, e.g., a single web service or app. (Automatically) combining the results from multiple such scans across many components of the same product in the light of a product-level risk model is a non-trivial task. Manually assessing scanner results as part of the development process, e.g., in a CI/CD pipeline, does not scale and leaves the risk of failing the conformity assessment before releasing the product.


\paragraph{Consistency}

The CRA mandates a risk model, which is used to determine how the essential requirements must be met by the product. Whether a certain measure is good enough depends on the identified risks. The severity of a vulnerability report that a manufacturer receives is also judged against this risk model. On the other hand, the risk model itself depends on the foreseen usage of the product and the environment in which it shall be operated. The CRA mandates that this information is documented and communicated to the user. Keeping all of these documents synchronized, not only between them, but also with the code especially in agile development processes, is a clear quest for novel apporoaches and tooling.

\paragraph{Lightweight Approaches}

The CRA is applicable to low-risk mobile games as well as high-risk industrial production machines. Development processes must therefore scale, while providing the same documentation and artefacts required for the CRA. The challenge is to keep the effort and workload for providing the risk model, the usage instructions, etc. and for conducting the conformity assessment as low as possible. Applying the same processes already in use, e.g., for medical software, would render the business model of an app infeasible.

\paragraph{Vulnerability Detection}

While various code scanning techniques have been proposed~\cite{arzt2014flowdroid,jovanovic2006pixy,8399530}, these approaches often rely on an implicit security model that is disconnected from the explicit risk model required by the CRA. Further, these techniques often focus on individual components such as web applications or device firmware. If a vulnerability is detected, it remains unclear which effect this component-level vulnerability has on the overall system given the risk model.

\paragraph{Software Bill of Materials} 

The CRA requires manufacturers to provide a software bill of materials (SBOM), i.e., the list of all first-level dependencies of the product. For simple software products, the SBOM can easily be constructed from Maven or Gradle definition files. Industrial control systems, on the other hand, may consist of multiple computers running different software, each of which relies on specific hardware drivers. These dependencies cannot easily be captured using, e.g., Maven. Existing approaches either require manual bookkeeping or rely on heuristics to match library signatures on existing code. Both approaches are insufficient on a large scale, especially in agile development processes.

	\section{Conclusion}
	\label{sec:conclusion}
	In this paper, we have highlighted the challenges that current legislation approaches in the EU have on software engineering as a research area, but also as an engineering discipline. As the EU is a significant market, many products are immediately affected by these challenges. Further, we assume that the EU's regulation can serve as a blueprint for other jurisdictions in the future, with security already being under explicit scrutiny around the world.

In the future, we will participate in addressing the challenges we have highlighted in this paper, with a special focus on automated code analysis techniques.

	\begin{acks}
		This research work has been funded by the German Federal Ministry of Education and Research and the Hessian Ministry of Higher Education, Research, Science and the Arts within their joint support of the National Research Center for Applied Cybersecurity ATHENE.
	\end{acks}

	\bibliographystyle{ACM-Reference-Format}
	\bibliography{paper}


\end{document}